\documentclass[sigconf, screen]{acmart}

\AtBeginDocument{%
  }

\setcopyright{none}
\renewcommand\footnotetextcopyrightpermission[1]{}

\acmConference[NordiCHI '26]{Nordic Conference on Human-Computer
  Interaction}{October 5--7, 2026}{Vaasa, Finland}
\acmYear{2026}

\usepackage{csquotes}
\usepackage{lscape}

\begin{document}

\title[Editorial Alignment]{Editorial Alignment: A Participatory
  Approach to Engaging Editorial Expertise in LLM-mediated Knowledge
  Dissemination}

\author{Simon Aagaard Enni}
\affiliation{%
  \institution{Aarhus University}
  \city{Aarhus}
  \country{Denmark}}
\email{enni@cas.au.dk}

\author{Malthe Stavning Erslev}
\affiliation{%
  \institution{Aarhus University}
  \city{Aarhus}
  \country{Denmark}}
\email{stavning@cc.au.dk}

\author{Karl-Emil Kjær Bilstrup}
\affiliation{%
  \institution{University of Copenhagen}
  \city{Copenhagen}
  \country{Denmark}}
\email{keb@di.ku.dk}

\author{Kristoffer Laigaard Nielbo}
\affiliation{%
  \institution{Aarhus University}
  \city{Aarhus}
  \country{Denmark}}
\email{kln@cas.au.dk}

\begin{abstract}
    The emergence of LLM-driven information services is reshaping the
    conditions under which public knowledge institutions operate,
    threatening to absorb the editorial function these institutions
    exist to exercise. While LLMs offer powerful new affordances for
    knowledge dissemination, editorial authority is challenged by
    pretrained LLMs that arrive already aligned with the values and
    dissemination strategies of their commercial developers. This paper
    investigates editor participation in re-aligning LLM interfaces to
    editorial standards through design workshops, in a case study where
    we design and implement an LLM-enabled encyclopedia interface with a
    Nordic public knowledge institution. We introduce \textit{editorial
    alignment} as a design practice within Participatory AI, framing AI
    alignment as a design process and positioning the editorial standard
    as a design artefact that translates editorial practice and values
    into alignment objectives for technical implementation. Last, we
    discuss how editorial alignment can create space for ongoing
    participation and give editors agency in LLM-mediated knowledge
    dissemination.
\end{abstract}



\maketitle
\section{Introduction}

The shift towards large language model (LLM)-driven information services is transforming the conditions under which knowledge institutions operate. Search engines like Google have long shaped \textit{what} information users encounter, through commercial ranking algorithms, yet the recent shift to a so-called ``zero-click-search'' paradigm \cite{Whi25}, where search results are presented through LLM-driven synthesis and summarization, extends this influence to \textit{how} the information is framed, contextualized, and engaged with as well \cite{ChaLie25}. This represents a structural shift in the online information landscape, as the editorial function of mediating between knowledge and a public, which is traditionally exercised by human professionals within accountable institutions, is increasingly being absorbed by models trained and governed by a small number of technology companies. For the organizations that constitute the public knowledge infrastructure of Nordic societies, such as national encyclopedias, public service broadcasters, libraries, and archives, this shift threatens their continued existence as intellectual authorities and caretakers of responsible knowledge dissemination.

At the same time, LLMs are powerful new tools for knowledge dissemination. Their capacity to adapt tone, vocabulary, structure, modality, and perspective in response to individual users' needs holds real promise for making public knowledge more accessible to those not well served by its current presentation. Students in particular are among those most rapidly embracing conversational AI interfaces for their knowledge needs \cite{Eva26}, sometimes even encouraged by their educational institutions \cite{Gueetal25}, and the public knowledge institutions targeting this demographic have to find ways to remain relevant as this technology reshapes user experiences. In the Nordic context, some institutions have chosen to embrace or deploy their own AI solutions, such as \textit{SMK - National Gallery of Denmark} through their SMK OPEN platform \cite{SMK_AI}, \textit{The National Encyclopedia} in Sweden through their AI-assistant AI-Martin\footnote{\url{https://www.ne.se/info/skola/ai-assistent/} - Accessed 22-04-2026} and \textit{Ordbogen A/S} through their \textit{Ordbogen AI} platform\footnote{\url{https://www.ordbogen.ai/} - Accessed 22-04-2026}, while others remain more careful, such as \textit{Lex---The Danish National Encyclopedia} and \textit{Store Norske Leksikon}, who have both explicitly banned the publication of AI-generated content \cite{SNL_KI, LEX_AI}.

For public knowledge institutions, generative AI presents both an existential threat and a transformative opportunity---yet realizing this potential without undermining the institutions' \textit{raison d'\^etre} as gatekeepers and guarantors of trustworthy knowledge can be difficult in practice. On the one hand, building a competitive LLM from scratch is technically and computationally prohibitive for all but the most well-resourced teams \cite{LuiDen21}. On the other hand, deploying pretrained LLMs carries its own risks, as such models are already aligned, at least partially, with the values and dissemination strategies of other organizations, notably the tech companies that create and maintain them~\cite{Wang_aligning_LLMs, johnson_ghost_in_the_machine}, which potentially compromises the editorial authority of the institution that deploys them. Any institution building on these foundations must therefore pursue the deliberate realignment of the system to its own institutional context. We use alignment here to refer broadly to the ensemble of processes and techniques used by designers and developers to steer and manage the inputs and outputs to an LLM-enabled system in order to reflect specific values, constraints, and conditions~\cite{Gab20}. In \autoref{sec:alignment_design}, we elaborate on this process and argue it is as much a design challenge as a technical one.

We engage this challenge through a case study with a Nordic online encyclopedia operating under precisely these institutional pressures. Designing an LLM-enabled interface in this setting led us to confront a prevalent dynamic in LLM integration efforts: the tendency to conceptualize such work as a rupture and a technological displacement of existing professional practice with something new and fundamentally different. Drawing on science and technology studies \cite{Agr97, Seletal19} and participatory design traditions~\cite{Bodetal21}, we instead pursue LLM integration as a process of negotiation with existing sociotechnical arrangements, aiming to engage with rather than replace the institutional practices and infrastructures already in place. In this view, the introduction of new technology is not the final output, but a first step for technological and professional development. In the case study, we center the professional practice of editors in the integration of LLM technologies, as the editorial work is the foundation of the encyclopedia's intellectual authority---it is what makes encyclopedic knowledge trustworthy and what distinguishes it from other sources of knowledge. It is also the editorial role that is perhaps most directly threatened by the shift to LLM-mediated knowledge dissemination. For these reasons, we choose this practice as a primary locus through which to align the system to the institution's values: both to secure the trustworthiness of the LLM-generated output, and as a commitment to giving the editors genuine agency in shaping how their institution's knowledge is disseminated. We term this approach \textit{editorial alignment}.

Editorial alignment is grounded in the emerging tradition of participatory AI \cite{Elmetal25}. A defining conceptual move in this paper is reframing AI alignment as a collaborative, practice-centered design activity conducted with and by the practitioners who embody the institutional values the system should reflect, rather than a technical optimization problem solved iteratively by system developers. This reframing establishes a methodological bridge between the AI alignment literature and participatory design, and challenges the often universalist and essentialist conceptions of ``human values'' common in the alignment literature by grounding value specification in the concrete, contextually specific, and critically examined practice of a real institutional community. 

While situated in a larger, longitudinal partnership and project, the case study is centered around two workshops with the editorial team of the institution. The first was a Future Workshop~\cite{Hansen2020_How_PD_works}, exploring editors' perspectives on AI-mediated knowledge dissemination, and the second was a practice-centered workshop, in which editors worked with LLM-generated text, individually in advance and collectively in session, to surface, discuss, and codify the tacit standards and latent values embedded in their professional work. This process resulted in an ``editorial standard'': a concrete design document specifying the constraints and conditions that LLM-generated output should aim to satisfy to uphold the institution's trustworthiness. This standard is envisioned as a living document, subject to ongoing editorial revision as both the system and the editorial practice evolve---and as an anchor for editorial participation in the governance of the LLM system.

The contributions of this paper are: 
(1) the introduction of editorial alignment as a design practice for LLM-based interfaces in editorially governed institutions; (2) a case study in a Nordic public knowledge institution, investigating how editors can participate in developing and governing AI alignment objectives; and (3) a discussion of how participatory AI approaches can position editors as resourceful and accountable actors in an organizational AI transformation.  

\section{Background}

This section situates the paper's contribution across three bodies of work: alignment as a design problem, participatory approaches to AI development, and the institutional landscape in which these concerns have become urgent. We begin by examining AI alignment as a design challenge rather than a problem of value specification, and introduce the alignment techniques available for LLM-based systems whose affordances and constraints define what can be meaningfully pursued as an alignment objective. From this we draw the methodological consequence that the design of aligned LLM systems for institutionally governed contexts requires participatory approaches that can surface professionally embedded values and translate them into alignment objectives. We situate this work in the traditions of participatory design and participatory AI. Finally, we turn to the institutional landscape in which this challenge is currently unfolding, surveying how Nordic public knowledge institutions have responded to LLM integration and identifying the structural gap in existing responses that this paper addresses.

\subsection{Alignment as Design}
\label{sec:alignment_design}

The concept of AI alignment has a long history as both a cultural and academic exercise. As a cultural idea, it goes back at least as far as 1940's sci-fi and Isaac Asimov's three laws of robotics \cite{Asi42}. The discussion has often been tied to a conception of AI that affords machines a degree of unpredictable agency and autonomy independent of human involvement \cite{Wie60, Rus19}, and value alignment, that is, aligning the values and objectives of AI systems with those of their human designers, users, or owners, is commonly proposed as a way to govern this autonomy \cite{Rus19, Gab20}. This conception presupposes that AI systems are capable of genuine autonomy and that their agency can be anchored in internalized values. Identifying such values is often framed as an exercise in ethics and morality, though prescribing moral behavior with sufficient detail and unambiguity is widely recognized as extraordinarily difficult \cite{Gab20}. Instead, it has recently become common practice to select a few broad ethical principles as a foundation for alignment---typically variations on helpfulness, harmlessness, and honesty---and supplement these with quantified expressions of ``human preference'' extracted from collected human feedback \cite{Chretal17, Baietal22, Linetal25}. In their seminal paper on Constitutional AI, \citeauthor{Baietal22} write: ``Our goal is not to define or prescribe what ‘helpful’ and ‘harmless’ mean but to evaluate the effectiveness of our training techniques, so for the most part we simply let our crowdworkers interpret these concepts as they see fit''~\cite[p. 4]{Baietal22}. Broad principles, in practice, are thus delegated to the interpretative discretion of individual annotators---grounding alignment less in ethical reasoning than in a kind of ``aggregated intuition.''

This practice relies on a universalist and essentialist conception of values \cite{Arzetal24}. Universalist, because the values applied are assumed to be either broadly applicable or broadly representative of universal human or cultural \cite{Pawetal25} values. Essentialist, in that human feedback is assumed to reveal essential human or cultural values and preferences that can be used as a ``ground truth'' for alignment through machine learning (ML). While there have been attempts at democratizing the formulation of alignment objectives through broad participation, in practice this often amounts to ``sourcing'' values and principles from a segment of a population and applying mathematical techniques to aggregate the results \cite{Huaetal24, MoaGan25}.

An alternative interpretation of AI alignment, which we employ in this paper, views ML models---the core of modern AI systems---as design artifacts inherently embedded in existing sociotechnical systems \cite{Seletal19, EnnAss21, Arzetal26, Elmetal25, Linetal25}. In this view, values are embedded in the models and in the larger systems in which they are used through a process of design, and the behavior of the models themselves cannot be disentangled from the sociotechnical context of their creation, implementation, and use. Design is strongly related to the notion of \textit{wicked problems}, that is, problems that are not clearly defined and that cannot be unilaterally solved \cite{Buc92}. This means that design necessarily entails bespoke prioritization and deliberation between multiple possible solutions. Rather than assuming that essential values can be sourced from human feedback or codified as universal ethical principles, a design-centric approach views AI alignment as an exercise in managing and implementing multiple, potentially contradictory, perspectives on the promise and potential of the designed system. Recent pragmatic turns to address the immense ethical implications and complexity of modern AI technologies come to similar conclusions---in both HCI~\cite{oz2026_turn_to_practice_HCI} and AI research~\cite{nyrup2022explanatory}. Further, this framing echoes the ways in which participatory design for decades has approached general technological development and integration projects~\cite{Bodetal21, Karasto2024_PD_infrastructuring}---a tradition we return to in the next section.

In this paper, we focus specifically on LLMs: systems that generate outputs by applying statistical regularities learned from large text corpora, rather than by executing explicitly specified symbolic logic. Whereas conventional software expresses desired behavior as deterministic rules that can be specified, verified, and adjusted directly, LLM behavior is evaluated empirically and shaped through a range of techniques. These include approaches that manipulate the model's input context \cite{Meietal25}, modules that monitor and filter outputs \cite{Donetal25}, and post-training strategies such as supervised fine-tuning or reinforcement learning from human feedback~\cite{Jietal23}. Typically applied in combination, these techniques each carry specific affordances and constraints that determine what kinds of alignment objectives can be meaningfully expressed through them. Formulating alignment objectives for an LLM-based system is accordingly an exercise in working within and across these constraints---closer to design than to specification, and one that places real demands on what a participatory alignment process can realistically aim to produce: objectives must be simultaneously grounded in situated professional values, expressible through available techniques, and amenable to empirical rather than formal evaluation.

\subsection{Participatory AI}
\label{sec:participatory_AI}

The characterization of alignment as a wicked design problem carries a direct methodological implication. Wicked problems cannot be resolved unilaterally, but require bespoke deliberation among multiple parties who bring different and partially incommensurable perspectives on what a system should do and who it should serve. For LLM-mediated knowledge dissemination, the parties relevant to this deliberation are not only technical, that is, developers and researchers who understand what alignment techniques can and cannot express, but also professional and institutional: the practitioners who embody the values and standards the system is meant to reflect, and who bear accountability for the quality of the knowledge it disseminates. Formulating alignment objectives that function simultaneously as accurate and situated descriptions of institutional values and as technically operationalizable constraints requires deliberation between these parties rather than unilateral specification by any one of them. Mediating between these different perspectives---technical, editorial, institutional---and their respective judgments means that the design of LLM-based interfaces for public knowledge institutions is well-suited for participatory design.

Thus, we situate this work in the traditions of participatory design (PD)~\cite{Bodetal21, Karasto2024_PD_infrastructuring} and the emerging field of participatory AI (PAI) \cite{Elmetal25}. PD research pioneered the participatory involvement of stakeholders and domain experts in the development and integration of sociotechnical systems, and has shown how this can create deep contextual knowledge and empower those affected by digital technologies~\cite{bodker2000co}. In this context, \textit{participatory} means that users, designers, management, researchers, and others collaborate towards shared goals and negotiate system designs throughout the design process. PD revolves around a \textit{third space} \cite{Mul02} that belongs neither to software developers nor to end users---a space where various stakeholders work together and learn from one another. In practice, this often takes place in workshop settings that frame the third space and seek to suspend the usual hierarchies of decision-making in favor of cross-cutting collaboration. In PD, people are viewed as fundamentally resourceful experts in their own domains, and the point of PD workshops is to empower them to put that expertise into play in the design of systems that have nontrivial impact on their professional lives. The role of the researcher in such workshops is that of the facilitator: guiding the process, helping stakeholders negotiate the situation, and participating actively rather than observing \cite{SanSta08}. PAI extends these commitments to the specific context of AI systems built on top of foundation models---systems that are ``fine-tuned, specialized, or otherwise adapted for specific domains and communities of practices'' \cite{Elmetal25}. PAI applies central PD principles to AI development with the aim of treating such systems ``as shared socio-technical systems that enhance rather than diminish human agency, human dignity, and human values'' \cite{Elmetal25}. Accordingly, PAI warrants that systems should be ``co-designed and fine-tuned with practitioners who understand domain-nuances, edge-cases and the social meaning of 'good' performance,'' and that the result should be ``locally-controlled AI systems that preserve cultural diversity and community self-determination'' \cite{Elmetal25}.

The application of PAI to LLM-based systems raises (at least) two challenges. The first concerns what participation means in the context of AI's data-driven development practices. As \citeauthor{Biretal22} observe, participation in AI development risks being reduced to the sourcing of data: human inputs are collected, aggregated statistically, and used to steer model behavior---with individual participants functioning as data generators rather than as deliberate co-authors of design decisions~\cite{Biretal22}. The pervasive application of post-training alignment techniques described above---RLHF, DPO, and related approaches---compounds this risk, as these methods tend to extract preference signals from human feedback and operationalize them through machine learning rather than through the kind of negotiated deliberation that PD envisions, thereby relegating participants to the role of data sources in the service of scalability. The distinction between these two modes of participation matters for what can be produced. If participation ends at ``sourcing'' values and preferences, participants' perspectives are flattened into aggregate patterns that are operationalized and interpreted in ways that leave little agency for those who produced them. If participation extends into the broader design and deliberation of the system, however, these same data---and any other design artifact produced by collective professional judgment---can be examined, contested, and revised by those who produced it. In the context of aligning an LLM to an institution's editorial standards, the latter makes the editorial team genuine agents in the alignment process rather than a source to be mined. The second challenge is that the sociotechnical complexity of modern LLMs makes genuine participation difficult to sustain in practice: when participants lack sufficient understanding of what a system does and how its behavior is governed, their responses tend toward the reactive rather than the evaluative, which is precisely the condition that makes data sourcing a tempting substitute for deliberation~\cite{Deletal23, CorDenEre23}. Developing participants' understanding of the system's possibilities and limitations is accordingly a precondition for meaningful participation in alignment work, which directly echoes traditional PD emphasis on empowerment of the participating stakeholders~\cite{ehn1988work}.

The data-sourcing risk is compounded when the values a participatory alignment process must elicit are themselves embedded in professional practice rather than codified in organizational policy. Much of what practitioners know about the standards of their field is tacit, that is, reliably exercised in practice, but resistant to full verbalization~\cite{Pol66}. \citeauthor{Sch83}'s account of professional expertise as grounded in \textit{knowing-in-action} and \textit{reflection-in-action} rather than the application of explicit rules captures this well: an experienced editor may immediately recognize a response that violates their standards, yet be unable to articulate in advance the principle by which they did so~\cite{Sch83}. The standards that constitute editorial authority are exercised through practice rather than derived from it in any straightforward way. However, this tacit knowledge is not a stable property of individual practitioners available for extraction, but is shaped by and distributed across the community of practice in which it is embedded, making it dynamic, resistant to extraction, and highly contingent on the specific social and institutional context in which that community operates~\cite{Mar12}. The implication for value elicitation is significant: asking practitioners to describe their standards in the abstract tends to surface either official policy accounts or personal rationalizations---neither of which captures the collectively negotiated, practice-embedded knowledge that actually governs professional judgment in context. PD's emphasis on situating deliberation in something that resembles actual work practice speaks directly to this kind of difficulty. Workshop formats that ask practitioners to exercise their judgment on concrete materials make tacit knowledge available in the form in which it is most reliable.

These core commitments of PD are reflected in a growing body of PAI case work. \citeauthor{Bilstrup2025_automation_integration} investigate how PD can be applied to the question of LLMs in K-12 education, focusing on how teachers can draw on their didactic competencies to meaningfully integrate language model technologies in their classrooms beyond mere task automation~\cite{Bilstrup2025_automation_integration}. In the domain of the creative professions, \citeauthor{Inie_2023_PAI_creative} survey how creative professionals view the potential impact of LLMs on their work, arguing that designing LLM-based systems for the creative domain should incorporate PAI approaches that attend to how practitioners may understand, cope with, adapt to, and exploit emergent technology~\cite{Inie_2023_PAI_creative}. A common thread runs across these cases: meaningful LLM integration depends on grounding the design process in the specific professional practice of the community the system is to serve. The relevant values and standards cannot be abstracted from the domain and treated as generic constraints applicable by any designer. Rather, they arise from, and are exercised through, the particular forms of work that practitioners do. PAI's central methodological commitment is to make this domain-specificity constitutive of the design process rather than a variable to be accommodated after the fact.

What the existing PAI literature has addressed less directly, however, is participatory involvement in AI alignment specifically. The cases surveyed above are concerned with how practitioners can participate in LLM integration decisions---how to adopt, configure, and evaluate tools in relation to their professional context. This is distinct from the question of how the LLM's behavior itself can be brought into alignment with the professional standards that practitioners embody, such that those standards govern system behavior persistently rather than being negotiated anew with each use. The distinction is sharpest in settings where the LLM operates as a public-facing interface that speaks on behalf of an institution: here, participation in integration decisions is insufficient unless it extends to the alignment objectives that determine what the system does when no practitioner is present to intervene. \citeauthor{Arzetal26} have moved in this direction, arguing that alignment should be grounded in situated norms and the concrete contexts of use in which misalignment actually emerges, rather than in abstract up-front specification~\cite{Arzetal26}. The present work builds on this to ask how a participatory process can be structured so that the tacit, practice-embedded character of professional standards---of the kind described above---can be both surfaced and translated into technically operationalizable alignment objectives. This is the specific challenge that the editorial context makes visible, and that editorial alignment is designed to address. In what follows, we develop this as a concrete design practice for editorially governed public knowledge institutions, drawing on a case study that tested its premises in practice.

\subsection{AI Integration in Nordic Public Knowledge Institutions}
\label{sec:ai_integration}

Nordic public knowledge institutions share a distinctive institutional profile: they position themselves as trustworthy alternatives to commercial information sources, maintain professional editorial staff accountable for the quality and integrity of their output, and ground their public legitimacy on this editorial authority. The recent proliferation of LLM-driven information services has forced these institutions to develop positions on AI integration, and examining the range of responses is instructive---not primarily as a survey of current practice, but because it reveals a structural gap that the field has not yet resolved.

The most prevalent and clearly articulated institutional response has been the establishment of boundaries around AI use in content \textit{creation}. Wikipedia's English-language community voted overwhelmingly (44 to 2) in March 2026 to prohibit the use of LLMs to generate or rewrite article content, citing hallucination, fabricated citations, the asymmetric burden of cleaning up AI-generated material relative to producing it, and the risk that AI-generated volume overwhelms volunteer review capacity \cite{Wik26}. The policy's framing reveals that proponents argued that Wikipedia's core product is not fluent text but \textit{traceable, source-grounded knowledge}---and that LLMs, optimized for plausibility rather than epistemic certainty, are structurally incompatible with this. \textit{Store Norske Leksikon} (SNL) has adopted an analogous position, banning AI-generated content on the grounds that language models are not truth-seeking systems: they predict probable word combinations rather than verify reality against primary sources, and thus cannot ground encyclopedic knowledge in the way that editorial accountability requires~\cite{SNL_KI}. The Danish National Encyclopedia, Lex, has published a detailed AI policy that similarly prohibits AI-generated content while explicitly permitting AI for internal workflows under strict conditions of human review~\cite{LEX_AI}. These positions converge on a shared logic: the institutional trustworthiness that distinguishes these encyclopedias from commercial information services depends on human editorial accountability, and AI-generated content threatens that accountability in ways that cannot be adequately governed through ex-post review.

The policies surveyed above do not, however, relate directly to content \textit{dissemination}. They govern whether the underlying encyclopedia articles are human-authored, not how those articles are synthesized, interpreted, and presented to users through an LLM interface. Yet it is precisely at the point of dissemination that LLM integration is proceeding most rapidly, and where editorial standards are most at risk of being bypassed without deliberate governance. Sweden's Nationalencyklopedin has integrated AI-Martin, an LLM-based assistant, directly into its encyclopedia platform for use in Swedish schools and the broader public~\cite{NE_AI}, without any publicly available editorial alignment framework governing how the system represents encyclopedic content, handles contested topics, or reflects Nationalencyklopedin's editorial standards.\footnote{Nationalencyklopedin is understood to be in the process of developing an editorial AI policy at the time of writing: the absence of a published framework reflects the rapidly evolving institutional situation rather than a lack of concern.} \textit{Encyclopaedia Britannica} has similarly deployed an AI chatbot interface for public use\footnote{\url{https://www.britannica.com/about-britannica-ai} - Accessed 22-04-2026} while maintaining restrictions on AI-generated articles, again without published alignment guidelines governing the dissemination interface. Interestingly, Store Norske Leksikon (SNL)---despite its principled position on content creation---itself experimented with building a proprietary AI chatbot, ultimately declining to launch because current models ``hallucinate too much'' for the institution to be willing to stand as the named responsible party for the output~\cite{Vaa23}. It is worth noting that this objection is not to AI dissemination in principle, but to AI dissemination without a credible means of aligning the system behavior with the editorial standards for which the institution is accountable.

The institutional landscape thus reveals a structural asymmetry: most Nordic public knowledge institutions have developed principled, well-reasoned positions on AI for content creation without any equivalent framework for AI-mediated dissemination. It is not the case that the former can simply be extended to the latter. The latter challenge requires not only deciding \textit{whether} to deploy an LLM interface, but \textit{how} to deploy it in a way that preserves rather than undermines the editorial authority on which the institution's trustworthiness rests. This is a challenge the encyclopedia sector has not yet systematically engaged.


Although not directly situated in the context of knowledge institutions such as encyclopedias, it is worth noting the relatively longstanding engagement with editorial work in an AI-context within Nordic journalism. Nordic newsrooms have been early and engaged adopters of AI tools for editorial workflows, motivated in part by a deliberate effort to maintain independence from the technology platforms that captured commercial value during earlier waves of digital transformation~\cite{Linetal24}. This has generated a body of practice and reflection on the relationship between AI capabilities and journalistic values~\cite{Ada25}. \citeauthor{Kometal20} argue that AI alignment in the newsroom must be grounded in \textit{situated} journalistic values rather than abstract ethical axioms---that what ``accuracy'' or ``fairness'' means can only be determined within the concrete professional context of a specific editorial community~\cite{Kometal20}. A recent synthesis of newsroom AI policies similarly finds that while existing guidelines prioritize transparency and human supervision, they are ``ill equipped to address subtle biases that may be built into third-party tools'' and rarely provide practical guidance for governing third-party LLM systems specifically~\cite{CNTI25}. The structural problem identified here---a gap between principled values and their operationalization in relation to external AI tools---is analogous to the challenge that encyclopedic institutions now face in the dissemination context.

\section{Editorial alignment}

We coin the term \textit{editorial alignment} to describe a design practice within participatory AI, applied to the specific setting of editorially governed institutions that introduce LLM-based interfaces for public knowledge dissemination. Classical editorial oversight---the sequential review of individual outputs by qualified editors before publication---is not a practical way to govern the quality of these outputs when texts are generated on demand in response to individual user queries. The institution therefore becomes dependent on the behavior of the generative model and the system surrounding it being aligned with the editorial standards and values that would otherwise be enforced through human review. The alignment target is consequently not abstract ethical maxims or generalized user preferences, but a situated, contested, and evolving body of professional practice representative of the editorial culture of the institution itself.

Building on the PD and PAI frameworks introduced above, editorial alignment treats editorial culture not as something \textit{a priori} or axiomatic---that is, as a fixed set of rules to be extracted and encoded---but rather as something continuously negotiated and arising in practice, shaped by the push-and-pull between the current members of the institution, the continuity and inertia of historical practice and its traces in the material, and the shifting expectations of the external world. The design challenge of editorial alignment is accordingly twofold: 1) to engage editors as professional experts in the formulation of alignment objectives, centering the accountability and expertise that are constitutive of the institution's legitimacy; and 2) to do so in a way that accommodates the dynamic nature of editorial practice, establishing alignment not as a one-time specification but as an ongoing process subject to the same professional participation and revision as the editorial practice it reflects \cite{Arzetal26, Elmetal25, Biretal22}.

The editorial standard is the central design artifact of editorial alignment: a document that captures important editorial principles and values in a form that can inform the implementation of alignment techniques. As such, it functions as a boundary object \cite{StaGri89, Mul02} between two communities of practice with distinct epistemic requirements---the editorial team, for whom the standard must accurately represent the professional judgment and situated values that govern their work, and the technical team, for whom it must translate into actionable alignment objectives compatible with available techniques. The development of such a standard is complicated by the nature of the knowledge it must capture. As established in Section~\ref{sec:participatory_AI}, professional editorial standards are largely tacit in character: reliably exercised in practice but resistant to full articulation in the abstract~\cite{Pol66, Sch83, Mar12}. This has direct consequences for how the editorial standard can be developed: it cannot be derived from organizational policy documents or abstract consultation with editors, but requires situating practitioners within something sufficiently close to actual editorial work that professional judgment can be exercised on concrete material rather than only described from memory.

There is accordingly a nontrivial tension at the heart of the editorial standard: a standard grounded in tacit, practice-embedded knowledge is, by definition, difficult to make fully explicit---yet technical alignment requires sufficient explicitness to operationalize the standard through available techniques. Editorial alignment does not claim to resolve this tension, but instead embraces it: by centering editorial practice as the primary site of elicitation, professional judgment can be accessed in the form in which it is most available, and by producing a text-based set of principles through collective deliberation, the practice yields something that can function as alignment objectives. The result is a principled approximation grounded in the concrete exercise of editorial work, functioning as a starting point rather than a final specification. Editorial alignment accordingly treats the editorial standard as a living document, subject to ongoing revision as professional practice evolves, as the system develops, and as new alignment challenges surface through situated use \cite{Arzetal26}---an ongoing anchor for editorial participation in the governance of the LLM system.

Existing participatory approaches to AI alignment focus either on individual end-users co-constructing alignment through runtime interaction \cite{Arzetal26}, or on organizational governance structures for institutional AI ownership \cite{Tseetal25}. Neither addresses the challenge specific to editorially governed public institutions: how to elicit alignment objectives from the tacit professional knowledge embedded in editorial practice, such that they can be technically implemented to govern system behavior persistently across interactions. Unlike general-purpose AI systems, whose alignment target is individual user preference and which do not speak on behalf of any particular institutional voice, LLM interfaces in public knowledge institutions are expected---by users and editors alike---to represent an institution whose trustworthiness and intellectual legitimacy is fundamentally tied to the quality and integrity of its dissemination. Unlike organizational governance, which concerns \textit{who} controls and owns AI systems, the challenge here concerns the \textit{content} of alignment: specifically, how the professional judgment exercised daily by editors can be brought into the alignment process. The need for such situated, practice-grounded alignment has been most clearly articulated in journalism, where \citeauthor{Kometal20} demonstrate that newsroom alignment must be grounded in concrete professional values rather than abstract ethical axioms~\cite{Kometal20}. Editorial alignment builds directly on this to offer a framework for how such grounded alignment is developed in practice, while addressing the dimension other approaches leave open: the content of alignment itself. As such, it is applicable across any media-producing institution that deploys LLM-mediated interfaces---public service encyclopedias, journalism and media organizations, public broadcasters, cultural heritage institutions, and archives---wherever editorial standards are at risk of being bypassed at the point of dissemination without deliberate governance.

Editorial alignment is not a framework that was designed in the abstract and subsequently tested against an institutional case. Rather, it was developed through and in response to the concrete challenges of an ongoing design project with a specific institution, and its theoretical commitments reflect the pressures and insights that emerged from that practice. In the following case study, we report on our experiences from designing and implementing an LLM-enabled interface for encyclopedic knowledge dissemination grounded in the professional practice of their editorial team.

\section{Case Study: Editorial Alignment in Practice}

Editorial alignment is, above-all, an approach that needs to be actualized via design work rather than a set of abstract principles, and its specific instantiation will change depending on the context. To demonstrate this, we present a case study of a project involving a Nordic online encyclopedia, in which we put editorial alignment to work. The purpose of the larger project is to design LLM-mediated interfaces to improve the accessibility of the material stewarded by the encyclopedic editors and organization without compromising its trustworthiness and responsibility in disseminating that material. In other words, this is a case where the alignment of the LLM to the editorial standards of the encyclopedia is crucial. For this reason, editors were identified early as central stakeholders in the alignment of the LLM, and we sought to engage their expertise via participatory workshop formats.

We conducted two workshops with editors at the encyclopedia; the first was a Future Workshop~\cite{Hansen2020_How_PD_works}, motivated by the need to create an initial mutual understanding and shared vision for the project, and the second was a specially-developed workshop aimed at specifying a set of editorial standards that could inform the alignment of the LLM-mediated interface. Both workshops were conducted in the participants' native language and quotations included below have been translated into English by the authors. We first report main findings from the Future Workshop in brief, since it provided context and direction for the second workshop, which we then describe in more detail. \autoref{tab:workshops} provides an overview of the two workshops.

One of the authors facilitated and participated in both workshops, while another was only present for the second workshop. During the workshops, they made observational notes and audio recorded the common conversations which were later transcribed. Further, they collected artifacts produced in the workshops: this includes post-it notes with critiques, visions, and implementation proposals from the first workshop and LLM conversations annotated and edited by the editors, and different drafts of the editorial standard developed in the second workshop. After each workshop the authors wrote a note that summarized what happened in the workshop, the main themes discussed, and the main outputs. These notes, together with written representations of the collected artifacts, were subsequently shared with all workshop participants. Based on these data together with first-person experiences, we report on how editors and researchers shared knowledge and collaborated on designing an editorial standard.

Prior to the workshops, the project had established three foundational design commitments that determined the scope of the interface: (1) the system should be faithful to the source material, generating responses grounded in the encyclopedia's content rather than drawing on the broader knowledge of the underlying model; (2) it should be bounded, deferring to the user when a query falls outside the encyclopedic domain---whether in topic, type, or tone---rather than attempting to answer from general knowledge; and (3) its responses should be relevant, prioritizing the most directly applicable material from the knowledge base over exhaustive coverage. These commitments defined what kind of interface was being built before the question of how it should communicate arose. The editorial standard developed in Workshop 2 presupposes and operates within this scope: it governs the character and quality of responses the system produces, not the prior question of whether the system should respond at all.

\subsection{Workshop 1: Future Workshop}

To engage the existing editorial practice, we arranged an initial Future Workshop with four senior representatives from the editorial team in late 2025 as well as one project manager and one researcher present. The four editor participants were all involved in the internal training program for new editors in the institution, which made them especially suited for articulating and disseminating the current editorial line for this project. The workshop was recorded and transcribed, together with post-it notes containing the participants feedback in the three phases of the workshop. Our focus is mainly on the second workshop where we developed the editorial standard. Therefore, we do not cover the full progression of the workshop but stick to reporting the main conclusions that came out of it.

\begin{landscape}
\begin{table*}[]
\begin{tabular}{llll}
\textbf{} &
  \textbf{Goal} &
  \textbf{Activities} &
  \textbf{Data \& Output} \\ \hline
\begin{tabular}[c]{@{}l@{}}\textbf{Workshop 1}\\ \\ Winter 2025\\ 3 hours\\ 4 editors\\ 1 project manager\\ 1 researcher\end{tabular} &
  \begin{tabular}[c]{@{}l@{}}A Future workshop~\cite{Hansen2020_How_PD_works}  to create \\ mutual understanding between the \\ two project partners and form a \\ shared vision for the project.\end{tabular} &
  \begin{tabular}[c]{@{}l@{}} \\ 1. Critique system prompt and outputs \\ from prototype of LLM system.\\ \\ 2. Envision how the LLM outputs should\\ ideally look like for different user groups\\ and how they should be able to interact\\ with the content through the LLM-system.\\ \\ 3. Discuss how the three highest prioritized\\ ideas can be implemented in the prototype\\ an organizational practices. \\ \hphantom{invisible text} \end{tabular} &
  \begin{tabular}[c]{@{}l@{}} \\ \textbf{Data collected:}\\ Observation notes, transcribed audio \\ recordings, and post-it notes with  \\ critiques, visions, and implementations, \\ sorted into themes.\\ \\ \textbf{Output:}\\ A shared understanding of each partners'\\ competencies and role in the project.\\ And a common set of concerns with and\\ opportunities of LLM-mediated \\ knowledge dissemination. \\ \hphantom{invisible text} \end{tabular} \\ \hline
\begin{tabular}[c]{@{}l@{}}\textbf{Workshop 2}\\ \\ Spring 2026\\ 2.5 hours\\ 4 editors\\ 1 project manager\\ 2 researchers\end{tabular} &
  \begin{tabular}[c]{@{}l@{}}An editorial alignment workshop \\ to capture editorial decisions and \\ rationales as these emerged via \\ editorial work with the goal of\\ creating  an actionable editorial \\ standard to guide the editorial \\ alignment.\end{tabular} &
  \begin{tabular}[c]{@{}l@{}} \\ 1. Critique, edit, and discuss multiple LLM\\ conversations according to the editorial \\ practices to create a list of transversal \\ values, best practices, and boundaries as a\\ draft of an editorial standard.\\ \\ 2. Analyze the draft to resolve or prioritize\\ contradictions and  ambiguities to the \\ degree that it is possible.\\ \\ 3. Apply the draft to hypothetical question-\\ answer pairs, designed to breach common \\ assumptions and explore edge cases.\\ \\ 4. Discuss and update the draft to agree on \\ an editorial standard that the project can\\ move forward with. \\ \hphantom{invisible text} \end{tabular} &
  \begin{tabular}[c]{@{}l@{}}\textbf{Data collected:}\\ Observation notes, transcribed audio\\ recordings, editors commented and\\ edited LLM conversations,  and the \\ three iterations of the drafted editorial \\ standard.\\ \\ \\ \textbf{Output:}\\ A draft of an editorial standard,\\ consisting of five core values and twelve\\ rules that can be found in Section \ref{sec:editorial_standard}.\end{tabular} \\ 
\end{tabular}
\caption{Overview of the two workshops in which we co-design an editorial standard for an LLM-mediated encyclopedia interface together with the responsible editors.}
    \label{tab:workshops}
\end{table*}
\end{landscape}

Participants identified source quality and reliability as the primary concern for implementation of an LLM interface to the encyclopedia. As one participant said, ``this concerns what [the encyclopedia] is. And [the encyclopedia] is a place where you can go to find reliable knowledge.'' The editors highlighted significant variation in article quality, outdated content, and internal contradictions across the source material. The concern was, in part, that the LLM might surface outdated or otherwise skewed material, which would not be fitting for its purposes. Although the encyclopedia does contain outdated articles, which the team is continuously working on updating, the editors were adamant that such material should not be surfaced on either the front page or the chatbot interface and, ideally, should only be findable if the user proactively searches for it.

A secondary concern related to audience mismatch. Whereas the intended audience for the encyclopedia was previously academically-trained professionals, today the encyclopedia is intended to be, as one participant put it, ``for everyone. It is for the broad, interested population.'' However, the initial prototype chatbot's outputs reflected the formal, encyclopedic tone of the underlying articles, which is poorly suited to the platform's primary user base of young people and non-academic users.

In response to these concerns, the workshop produced a set of action points, including the technical filtering and weighting of sources based on quality and recency and a redesign of the tone and style to better serve younger audiences, accounting for varying levels of literacy. In addition, a longer-term outcome was the identification of a need to develop a framework for source criticism for AI systems, proposed as a collaborative effort with secondary education institutions. However, these action points lacked concreteness in terms of editorial standards. It was clear that the editors had an idea about what kind of tone and style was fitting for the system, but it was less clear how to formalize that in a way that could be implemented into the system without needing editorial oversight of each individual output. To concretize, we conducted a second workshop focused on editorial standards with the ambition to elicit editors' expertise via a practice-centric approach.

\subsection{Workshop 2: Deliberating an Editorial Standard}

To arrive at an actionable editorial standard to guide the editorial alignment process, we conducted a second workshop in spring 2026 intended to capture editorial decisions and rationales as these emerged via editorial work, rather than abstract descriptions of the importance of factuality and nuance. For this workshop we again invited four senior editors (three overlapping from the first workshop), as well as the same project manager and two researchers present. Whereas the Future Workshop centered on ideation and critique of the prototype, the second workshop had a more focused aim: to derive a concrete set of editorial principles and values that could constitute a working editorial standard for the AI-generated text produced by the LLM interface. Crucially, this standard was not to be imposed externally, but elicited through the editors' own professional practice. These standards might be partially codified already, but are likely also embedded in the professional practice and working culture of the institution. The resulting editorial standard would ideally translate editorial practice and values into alignment objectives for technical implementation, although it, as we describe above, necessarily exists at the crux of diverging kinds of knowledge. At this crux, the workshop sought to base itself on processes that mirror or otherwise resemble the established editorial workflows of the team in order to identify important situated aspects of the editorial standard that emerged only as an editor encountered a specific kind of text and immediately knew what to do with it, despite not being aware of that knowledge beforehand.

The workshop entailed that editors engaged in editorial work both before the actual workshop session, as preparation, and during the session itself:

\begin{enumerate}
    \item In advance, participants are presented with 20 exemplar question-answer pairs from actual users' interactions with a public prototype version of the AI interface. Participants are asked to each select three examples to edit according to their own professional practice, as individual preparation.
    \item As the workshop begins, the participants each present their chosen examples and the changes and critique that arose in their editing work. Following each presentation, the participants and organizers work together to identify and discuss provisional values, best practices, and boundaries, which are then added to a dynamic list that is updated as the workshop progresses. The goal at this stage is to identify as many values, best practices, and boundaries as possible, even if some of them implicitly contradict one another.
    \item After the presentations, the list of values, best practices, and boundaries is analyzed and discussed with the entire group, and any contradictions or ambiguities are resolved via general discussion in the group. In the case that agreement is not attainable, some prioritization or other form of specification should be added so that it becomes clear when to prioritize each of the potentially-conflicting entries.
    \item After a list of values, best practices, and boundaries has stabilized, the participants are presented with a series of new, purpose-specific and hypothetical question-answer pairs that they are asked to critique and edit as a group. These question-answer pairs are specifically intended to breach common assumptions shared among the participants (identified as part of Workshop 1) in order to surface fringe-cases where the editors respond to unexpected yet not irrelevant kinds of exchanges. At this stage, the participants are met with the additional restriction that all edits should be grounded in one or more entries on the list of values, best practices, and boundaries. If an edit is deemed necessary but it cannot be supported by the list of values, best practices, and boundaries, the list should be updated either by editing existing points or appending new ones.
    \item Finally, the list of points is prioritized in a collective discussion according to their value and effort for implementation.
\end{enumerate}

\subsubsection{The Editorial Standard}
\label{sec:editorial_standard}

The workshop concluded with the consolidation of an editorial standard, organized around five core values:

\begin{itemize}
    \item Respect the reader's time and attention
    \item Provide appropriate context
    \item Present content pedagogically
    \item Do not talk down to the reader
    \item Maintain a neutral and measured tone
\end{itemize}

\noindent These values were operationalized into twelve concrete rules to guide implementation:

\begin{itemize}
    \item Open with a lead paragraph that summarizes the answer and key points
    \item Signal how the user's query has been interpreted
    \item Define essential concepts
    \item Minimize textual complexity, academic register, and unnecessary jargon
    \item Place historical facts in their geographic and chronological context
    \item Follow [the encyclopedia]'s existing style guidelines on formal criteria (e.g., on abbreviations)
    \item Use examples to illustrate key concepts, but only from the source material
    \item Reserve high detail and elaboration for the body text following the lead paragraph
    \item Avoid normative or emotional judgments
    \item Avoid directly addressing the reader
    \item Avoid making assumptions about the reader
    \item Avoid figurative or narrative language
\end{itemize}

The workshop's most immediate consequence was a restructuring of the system's output format on the basis of the editorial standard. Responses were reoriented away from a conversational style toward a more structured presentation: a third-person, impersonal format that opens with a lead paragraph summarizing the answer, followed by an optional elaboration of how the query was interpreted and a more detailed body section for readers who require it. These changes were implemented primarily through context engineering and constrained generation and were chosen due to their high priority for the participants and their high compatibility with available alignment techniques.

\subsection{Analysis of Workshop 2}

We found that the workshop structure sustained different kinds of editorial insights at different stages, depending on what kind of professional practice each stage harnessed. In the first part of the workshop (stages 2--3 described above), participants focused on editorial decisions related to content and precision, whereas in the second part (stage 4), focus shifted to tone and style. We elaborate on this difference below, which we take to indicate different forms of editorial knowledge that also underscore two different aspects of the encyclopedia's ethos, each of which are put under pressure by the introduction of LLM-based dissemination.

\subsubsection{Presentation of Pre-edited Examples}

In advance, participants had selected and editorially critiqued three responses each from a set of 20 real user interactions with the current prototype. In the first part of the workshop, participants presented their critiques and the editorial reflections they had generated.

Editors largely focused on the content of the responses and how they reflected the source texts. One example involved an overview of a specific dialect, which opened up a discussion of how much information the LLM response should include. Although there was general agreement that the output should not be excessively long, the presenting editor found it unsatisfactory that the response lacked explanation of central terms. In a related case, an editor noticed that a response related to a form of neurodivergence lacked important information that would introduce necessary nuance to avoid strengthening existing negative biases related to the neurodivergence in question. The issue was not that the information provided was untrue, but that it was insufficient to effectively combat the propagation of unwarranted negative bias. When asked about the consequence of an LLM output potentially worsening a negative bias in this way, the editors agreed that, as one participant put it, ``that would be very problematic because I think you have to consider the chatbot to be part of the encyclopedia.'' This same participant added that in such a case, the output ``will have to be more complex. It will have to be longer.''

That said, another significant concern of the editors when they presented their prepared edits related to precision in terms of structure. The editors agreed that it was problematic for the LLM responses to answer the queries sequentially and only present the conclusion at the end. In one extreme case, a user had asked the LLM about train departures for a specific urban commute, to which the LLM had first provided a geographical and historical account of the specified train stations and only \textit{then} highlighted that an encyclopedic chatbot system was ill-suited to answer such a query. Instead, an optimal structure would present a brief overall conclusion first---including, if relevant, rejection or redirection to other services---and only then provide a more in-depth answer below. The leading conclusion should also, at least for sufficiently complex queries, contain a disambiguation and contextualization of the topics covered. As one editor stressed, ``we contextualize anything we describe, whether it is a concept, a person, or an era, we always contextualize it chronologically and geographically.'' Following such a structure, leading with a conclusion containing some contextualization, would correspond to existing guidelines used for the underlying encyclopedic entries.

In addition, the editors commented on the relatively high linguistic complexity of the outputs, which largely corresponded to the tone of the underlying encyclopedic entries. The suggestion of a leading conclusion and contextualization was largely informed by a desire to appeal to readers who may not be comfortable reading longer outputs with higher levels of detail and less clear conclusions. However, as becomes evident below, the editors were largely in favor of conforming to a formal and somewhat academic tone and style, compared to more supposedly-engaging forms of dissemination.

\subsubsection{Reacting to Manufactured Edge-cases}
\label{sec:edge_cases}

In the second part of the workshop, participants evaluated four responses generated prior to the workshop by the two researchers participating. These manufactured responses were generated using Claude Sonnet 4.6 with prompts deliberately intended to push the boundaries of style and communicative register. The four prompts mimicked distinct personas: a plain-language factual assistant, a narrative popular-science communicator, a peer-register assistant aimed at middle-school students, and a social-media-style communicator using emojis and an attention-economy rhetoric. These responses were not grounded in the source material of the encyclopedia and, accordingly, focus was less on the correspondence between output and underlying material and more on tone and style. The persona prompts were written with the purpose of breaching existing linguistic practices and provoking boundary-setting. The linguistic complexity of the encyclopedia was a theme in both Workshop 1 and in the first part of Workshop 2, and the persona prompts were instrumental in specifying what kind of linguistic changes the editors would deem appropriate.

The exercise elicited strong reactions and surfaced several implicit principles that had not yet been articulated. Participants rejected any response that could be perceived as patronizing or condescending, employed figurative or narrative language, made assumptions about the reader's background or understanding, or adopted a conversational, ``chatty'' tone. Although the editors agreed that the manufactured responses were well-constructed in terms of what information---and how much information---they relayed, they also agreed that, as one participant put it, ``the tone is off somehow.'' Specifically, participants reacted to the use of first- and second-person pronouns that typically structure the chat format known from most LLM-based products. As one participant argued, ``(...) as soon as you start addressing the reader directly, it gets difficult, right? (...) when you start trying to crawl into the reader’s head, you’re on rather shaky ground, I think.'' This is a clear stylistic divergence from conventional implementation of LLM-based interfaces.

These reactions reflect a professional disposition that ran consistently through the workshop: as the party institutionally accountable for the quality of the dissemination, editors set limits on what the system may do, while it falls to other stakeholders---management, designers, and ultimately users---to push what is done within those limits. Although content may remain the primary concern, for good reason, the editors reacted almost viscerally to the edge cases as though the very identity of the encyclopedia was at stake. This indicates that the role of public service institutions is as much carried through style as content. Despite having an ambition to reach new audiences who do not intuitively connect to the encyclopedic tone of voice, the editors were prepared to sacrifice (some) readership in favor of preserving a tone that they thought was fitting. The notion that responsible dissemination might cause more work for the reader was acknowledged explicitly by one of the participants: ``We \textit{are} an encyclopedia, and it's okay for us to require our reader put in some work. That's okay.'' Of course, the editors’ opinions are not managerial in kind and may not correspond to the overall strategy of the encyclopedia at large, but they are indicative of what a public service institution is, editorially speaking. In this sense, the question of style and presentation becomes an indicator of the ethic of public service and the demands that such an ethic puts on the public that receives the service.

The use of persona prompts to surface edge cases resulted in the articulation of nontrivial editorial boundaries that the editors had not succeeded in verbalizing, or even recognizing, during either Workshop 1 or during the first part of Workshop 2, even though the question of style had come up on both occasions. In this way, the introduction of purposefully misaligned output was well-suited for activating tacit knowledge that was not available otherwise.

\section{Discussion}

One of the more striking outcomes of the editorial alignment workshops was the character of the editorial standard that emerged. When editors worked directly with LLM-generated text and collectively deliberated on what it should and should not do, the resulting standard was closer to the requirements governing encyclopedia articles proper than initially expected. No direct address, no assumptions about the reader's background or prior knowledge, no figurative or narrative language, no emotional register, no evaluative judgment unsupported by the source material: taken together, these principles stand in clear tension with the conversational affordances of the chatbot format that had been pursued up to that point in the project. While this could be read as a limitation of the method, or as evidence that editors failed to imaginatively engage the possibilities of the technology, we argue instead that it is a central finding: an indication that an editorial alignment process grounded in actual professional practice, rather than abstract consultation, surfaces genuine institutional values rather than socially desirable generalities.

This conservatism is not incidental to editorial practice but constitutive of it. The editors who participated are professionally accountable for the trustworthiness of the knowledge the institution disseminates, and their standards reflect an orientation toward responsibility over accessibility. As one participant put it, the encyclopedia is a place ``where you can go to find reliable knowledge''---and LLM-generated text that departs from these criteria, however accessible or engaging, risks undermining precisely the institutional authority that distinguishes the encyclopedia from other information sources. As mentioned in Section \ref{sec:edge_cases}, the editors were reluctant to address readers directly or attempt to ``crawl into the reader's head'' for exactly this reason. At the same time, conservatism is not the only value in play. While management regards the trustworthiness of the dissemination as essential, they place equal or perhaps even greater emphasis on accessibility, particularly for younger and less educated audiences, and are more willing to accept stylistic departures from the formal encyclopedic register in service of this goal. Rather than viewing this tension as a problem that must be resolved before design work can proceed, we suggest it is more accurately understood as the design work itself. Editorial alignment, as a practice, is an exercise in managing competing and partially incommensurable values---precisely the condition that defines design as distinct from specification~\cite{Buc92}. One productive way to frame this is to understand editorial practice as establishing the outer boundaries of responsible dissemination: the limits within which the system may operate without jeopardizing institutional trustworthiness. Within those boundaries, there is latitude for further design work---concerning, among other things, how the system addresses different audiences and navigates the tension between encyclopedic and conversational registers---that involves stakeholders beyond the editorial team. This framing situates the editors' contribution where it is most reliable, while acknowledging that it does not exhaust the design problem.

The editorial alignment process as conducted in this case study raises a serious issue that its commitment to PD makes it difficult to sidestep: does it actually escape the data-sourcing dynamic it critiques? The critique, developed in Section~\ref{sec:participatory_AI}, holds that participation in AI development risks being reduced to the extraction of human preferences, which are then operationalized by developers in ways that leave the original participants without genuine agency over the result~\cite{Biretal22}. The workshops described here represent a more substantive form of engagement: editors did not merely generate preference data but collectively deliberated on principles, contested each other's judgments, and produced a design artifact that carries the weight of their professional reasoning. The editorial standard will have real, concrete consequences for the system's implementation. And yet the project is owned by management, not the editorial team, and the editorial standard can be overridden or quietly de-prioritized without any formal requirement for editorial ratification. Editors were given influence in the design process---they were not given power over it. This leaves the fundamental asymmetry of the data-sourcing arrangement structurally intact, even when the quality of participation within that arrangement is more substantial. This does not invalidate editorial alignment as a practice, but it does identify where this instantiation of the practice falls short of its own aspirations, and what institutional changes that would be required to close this gap.

The question of who participates in editorial alignment, and why editors are centered specifically, warrants further reflection in light of this gap. The Scandinavian PD tradition, from which PAI draws its commitments, offers an instructive historical parallel in its early development. Here, focus was on the workers most directly affected by technological transformation---typesetters, in a paradigmatic newspaper case---on the grounds that their skills, their livelihoods, and their professional accountability were most immediately at stake~\cite{Bodetal21, Elmetal25}. This was, at its core, an epistemic and ethical choice rather than a pragmatic one: the workers who bore the accountability for the quality of the work were understood to be the parties whose participation was most consequential for the integrity of the resulting system. The same logic applies to editorial alignment. Editors are the practitioners professionally accountable for the quality and responsibility of the institution's knowledge dissemination---it is their judgment, exercised daily in practice, that constitutes the institution's intellectual authority, and it is their professional role that is most directly threatened by the shift to LLM-mediated dissemination. Centering their participation thus follows from the institutional logic of editorially governed public knowledge institutions and their democratic accountability. It is worth noting, however, that early Scandinavian PD's focus on workers also left other parties---readers, writers, and the broader public served by these institutions---largely outside the participatory process. Whether and how to extend participation in editorial alignment beyond the editorial team, toward the audiences the institution exists to serve, remains an open question that the present case study does not resolve---though one that the field of PAI seems well positioned to take up.

What would it mean for editorial alignment to close the gap between influence and power, and to provide editors with genuine ownership of the LLM system rather than a consultative role in its development? Our analysis suggests several implications. Most immediately, it would require that editors be given formal co-ownership of the project---a structural signal that their professional judgment carries equal institutional weight to management and research perspectives in decisions about the system's direction. More fundamentally, it would require reconceptualizing the editorial standard not merely as a design artifact produced in a workshop, but as something closer to a constitutional document: a set of principles that cannot be overridden without a formal editorial process, analogous to the existing standard maintained for the writing and editing of encyclopedic content, and that therefore serves as an anchor for ongoing editorial participation rather than a one-time contribution to a design project. Giving the editorial alignment workshop a recurring character---where editors periodically review AI-generated outputs both to evaluate compliance with the standard and to surface emerging misalignments---could be a way to establish a tight link between editorial practice and system alignment that serves two distinct purposes: as a technical mechanism for maintaining and improving system alignment, and as a structured, recurring form of professional participation that keeps the editorial team in an active and reflexive rather than merely historical relationship with the system they helped shape~\cite{Arzetal26}. Together, these arrangements would more fully realize what PAI recommends and what the institutional logic of editorially governed knowledge institutions demands: that the practitioners accountable for the quality of knowledge dissemination also govern the systems through which that dissemination increasingly occurs. Editorial alignment, as practiced in this case study, is a step toward that arrangement. Whether it arrives there depends less on the design practice itself than on the institutional will to restructure governance accordingly.

\section{Conclusion}

The shift toward LLM-mediated knowledge dissemination poses a structural challenge for public knowledge institutions whose authority rests on editorial accountability: how to integrate these technologies without ceding the editorial function that grounds the institution's trustworthiness. This paper has proposed editorial alignment as a design practice within participatory AI that addresses this challenge by treating AI alignment not as a technical optimization problem but as a collaborative, practice-centered design activity conducted with and by the editors who embody the institutional values the system should reflect. Through a case study with a Nordic online encyclopedia, we have shown how a participatory workshop process, grounded in concrete editorial work rather than abstract consultation, can surface tacit professional deliberation and translate it into an editorial standard that functions simultaneously as a representation of institutional values and as a set of alignment objectives for technical implementation. The case also surfaces an unresolved tension: giving editors influence in the design process does not, by itself, give them power over the system's governance, and without structural changes---formal co-ownership, a recurring editorial review process, and an editorial standard that carries constitutional rather than merely advisory status---editorial alignment is insufficient to substantially break with the data-sourcing dynamic that structured much participatory involvement in AI. The contribution of this paper is accordingly a design practice that editorially governed institutions can use to establish the boundaries of responsible LLM-mediated dissemination, with the recognition that realizing the full potential of this approach demands institutional arrangements in which editors are not only consulted but genuinely empowered.

\section{Generative AI usage disclosure}
Claude 4.6 Sonnet has been used to review, reword, restructure, and suggest draft text throughout the article. All text drafted or edited in this process has subsequently been reviewed, rewritten, and edited by one or more of the human authors.

\bibliographystyle{ACM-Reference-Format}
\bibliography{bibliography}
\end{document}